# Bibliometric-enhanced Retrieval Models for Big Scholarly Information Systems

Philipp Mayr
Knowledge Technologies for the Social Sciences
GESIS – Leibniz Institute for the Social Sciences
Cologne, Germany
philipp.mayr@gesis.org

Peter Mutschke
Knowledge Technologies for the Social Sciences
GESIS – Leibniz Institute for the Social Sciences
Cologne, Germany
peter.mutschke@gesis.org

*Abstract*— Bibliometric techniques are not yet widely used to enhance retrieval processes in digital libraries, although they offer value-added effects for users. In this paper we will explore how statistical modelling of scholarship, such as Bradfordizing or network analysis of coauthorship network, can improve retrieval services for specific communities, as well as for large, cross-domain large collections. This paper aims to raise awareness of the missing link between information retrieval (IR) and bibliometrics / scientometrics and to create a common ground for the incorporation of bibliometric-enhanced services into retrieval at the digital library interface.

## I. BACKGROUND

The information retrieval (IR) and bibliometrics / scientometrics communities move more closely together with combined recent workshops like "Computational Scientometrics" held at iConference 2013 and "Combining Bibliometrics and Information Retrieval" held in July at the ISSI conference 2013. During these previous workshops it became obvious that there is a growing awareness that exploring links between bibliometric techniques and IR could be beneficial for actual both communities. They also made visible that substantial future work in this direction depends from an awareness rise in both communities. This paper is a follow-up and shorter version of our recent paper in Scientometrics (Mutschke et al 2011) and is strongly associated with the DFG-funded research project "Value-added Services for Information Retrieval".

## II. INTRODUCTION

Traditional retrieval has reached a high level in terms of measures like precision and recall, but scientists and scholars still face challenges present since the early days of digital libraries: mismatches between search terms and indexing terms, overload from result sets that are too large and complex, and the drawbacks of text-based relevance rankings. Therefore we will focus on statistical modelling and corresponding visualizations of the evolving science system. Such analyses have revealed not only the fundamental laws of Bradford and Lotka, but also network structures and dynamic mechanisms in scientific production (Börner et al. 2011). Statistical models of scholarly activities are increasingly used to evaluate specialties, to forecast and discover research trends, and to shape science policy (Scharnhorst et al. 2012).

Introducing an IR perspective in science modeling is motivated by the fact that scholarly IR as a science of searching for scientific content can be also seen as a special scholarly activity that therefore should also be taken into account in science modeling. Moreover, as scholarly digital libraries (DLs) can be considered as particular representations of the science system, searching in DLs can be seen as a particular use case of interacting with exactly that system that is addressed by science modeling. From this perspective, IR can play the role of a validation model of the science models under study.

While, in this paper we mainly focus on how to use science model-enhanced IR as a test bed for different science models, we would also like to point out that there is a further interface between IR and scientometrics which is currently underexploited. One of the problem solving tasks shared by IR and scientometrics is the determination of a "proper" selected set of documents from an ensemble. In particular for newly emerging interdisciplinary fields and their evaluation the definition of the appropriate reference set of documents is important. Glänzel et al. (2009) have discussed how bibliometrics can be also used for the retrieval of "core literature". Bassecoulard et al. (2007) and Boyack & Klavans (2010) proposed sophisticated methods to delineate fields on the basis of articles as well as journals. However, due to the interconnectedness of research streams and different channels of knowledge transfer, it remains a complex problem how "hard boundaries" in continuously changing research landscapes can be found.

In their paper on Bibliometric Retrieval Glänzel et al. (2009) apply a combination of methods. They start from bibliographic coupling and keyword-based search and continue with a step-wise process to filter out the final core set from potentially relevant documents. Hereby, they make use of methods that are standard techniques in traditional IR as well (such as keyword-based search or thresholds). But, as already stated by Glänzel et al., "the objectives of subject delineation in the framework of bibliometric (domain) studies essentially differ from the goals of traditional information retrieval". In principle, this requires the application of different methods.

The bibliometric retrieval approach, in particular is an evaluative context, aims at defining a reference set of documents on the basis of a firm methodological canon, in order to justify the application and interpretation of standardized indicators. In traditional IR, in contrast, the application of bibliometric models and approaches has the primary goal to enhance the search from the perspective of the user by combining a wider search space with a particular contextualization of the search. The overall aim here is to help the user to get a grasp about the size and structure of the information space, rather than forcing him to precisely define the search space.

The models are presented in this paper therefore is to improve retrieval quality in scholarly information systems by computational science models that reason about structural properties of the science system under study.

### III. MODELS

Computational science models, to our understanding, are particular conceptualizations of scholarly activities and structures that can be expressed in algorithms (to be operationalized in systems that – more or less - reason about science, such as IR systems). The paper proposes two different kinds of science models as value-added search services that highlight different aspects of scholarly activity (see Figure 1): (1) a bibliometric model of re-ranking, called Bradfordizing, representing the publication form of research output and its organization on a meso-level in terms of journals (BRAD), and 2) a co-authorship model of re-ranking examining the collaboration between the human actors of knowledge flow in science (AUTH). Thus, the models address very different dimensions of structural properties in the science system. Moreover, they are also heterogeneous as regards the methods applied. BRAD uses bibliometric statistics, and AUTH methods taken from social network analysis, graph theory respectively. However, to the same extent as different science models emphasize different aspects of scholarly activity we expect that different kind of searches are best served by relying on corresponding science models.

#### A. Coreness of Journals

Journals play an important role in the scientific communication process. They appear periodically, they are topically focused, they have established standards of quality control and often they are involved in the academic gratification system. Metrics like the famous impact factor are aggregated on the journal level. In some disciplines journals are the main place for a scientific community to communicate and discuss new research results. These examples shall illustrate the impact journals bear in the context of science models. Modeling science or understanding the functioning of science has a lot to do with journals and journal publication characteristics. These journal publication characteristics are the point where Bradford law can contribute to the larger topic of science models.

Fundamentally, Bradford law states that literature on any scientific field or subject-specific topic scatters in a typical way. A core or nucleus with the highest concentration of papers - normally situated in a set of few so-called core journals - is followed by zones with loose concentrations of paper frequencies. The last zone covers the so-called periphery journals which are located in the model far distant from the core subject and normally contribute just one or two topically relevant papers in a defined period. Bradford law as a general law in informetrics can successfully be applied to most scientific disciplines, and especially in multidisciplinary scenarios (Mayr 2013).

Bradfordizing, originally described by White (1981), is a simple utilization of the Bradford law of scattering model which sorts/re-ranks a result set accordingly to the rank a journal gets in a Bradford distribution. The journals in a search result are ranked by the frequency of their listing in the result set, i.e. the number of articles in a certain journal. If a search result is "bradfordized", articles of core journals are ranked ahead of the journals which contain only an average number (Zone 2) or just few articles (Zone 3) on a topic (compare the example in Figure 1). This re-ranking method is interesting because it is a robust and quick way of sorting the central publication sources for any query to the top positions of a result set such that "the most productive, in terms of its yield of hits, is placed first; the second-most productive journal is second; and so on, down through the last rank of journals yielding only one hit apiece" (White 1981).

Thus, Bradfordizing is a model of science that is of particular relevance also for scholarly information systems due to its structuring ability and the possibility to reduce a large document set into a core and succeeding zones. On the other hand, modeling science into a core (producing something like coreness) and a periphery always runs the risk and critic of disregarding important developments outside the core.

#### B. Centrality of Authors

The background of author centrality as a network model of science is the perception of "science (as) a social institution where the production of scientific knowledge is embedded in collaborative networks of scientists" (He 2009). Those networks are seen as "one representation of the collective, self-organized emerging structures in science" (Börner and Scharnhorst 2009). Moreover, because of the increasing complexity of nowadays research issues collaboration is becoming more and more "one of the key concepts in current scientific research communication" (Jiang 2008).

Collaboration in science is mainly represented by co-authorships between two or more authors who write a publication together. Transferred to a whole community, co-authorships form a co-authorship network reflecting the overall collaboration structure of a community. Co-authorship networks have been intensively studied. Most of the studies, however, focus mainly either on general network properties (see Newman 2001, Barabasi et al. 2002) or on empirical investigation of particular networks (Yin et al. 2006, Liu et al. 2005). To our knowledge, Mutschke was among the first who pointed to the relationship between co-authorship networks and other scientific phenomena, such as cognitive structures (Mutschke 1994, Mutschke and Quan-

Haase 2001), and particular scientific activities, such as searching scholarly DLs (Mutschke 1994).

From the perspective of science modeling it is important to note that, as co-authorships also indicate the share of knowledge among authors, "a co-authorship network is as much a network depicting academic society as it is a network depicting the structure of our knowledge" (Newman 2004). A crucial effect of being embedded in a network is that "some individuals are more influential and visible than others as a result of their position in the network" (Yin et al. 2006). As a consequence, the structure of a network also affects the knowledge flow in the community and becomes therefore an important issue for science modeling as well as for IR (cp. Mutschke and Quan-Haase 2001, Jiang 2008, Lu and Feng 2009, Liu et al. 2005).

This perception of collaboration in science corresponds directly with the idea of structural centrality (Bavelas 1948; Freeman 1977) which characterizes centrality as a property of the strategic position of nodes within the relational structure of a network. Interestingly, collaboration in science is often characterized in terms that match a particular concept of centrality widely used in social network analysis, namely the betweenness centrality measure which evaluates the degree to which a node is positioned between others on shortest paths in the graph, i.e. the degree to which a node plays such an intermediary role for other pairs of nodes. Yin et al. (2006) see co-authorship as a "process in which knowledge flows among scientists". Chen et al. (2009) characterize "scientific discoveries as a brokerage process (which) unifies knowledge diffusion as an integral part of a collective information foraging process".

The betweenness-related role of collaboration in science was confirmed by a number of empirical studies. Yan and Ding (2009) discovered a high correlation between citation counts and the betweenness of authors in co-authorship networks. Liu et al (2005) discovered a strong correlation between program committee membership and betweenness in co-authorship networks. Mutschke and Quan-Haase (2001) observed a high correlation of betweenness in co-authorship networks and betweenness of the author's topics in keyword networks. High betweenness authors are therefore characterized as "pivot points of knowledge flow in the network" (Yin et al. 2006). They can be seen as the main driving forces not only for just bridging gaps between different communities but also, by bringing different authors together, for community making processes..

This strongly suggests the use of an author centrality model of science also for re-ranking in scholarly IR (cf. Zhou et al. 2007). The general assumption of the model is that a publication's impact can be quantified by the impact of their authors which is given by their centrality in co-authorship networks. Accordingly, an index of betweenness of authors in a co-authorship network is seen as an index of the relevance of the authors for the domain in question and is therefore used for re-ranking, i.e. , a retrieved set of publications is re-ranked according to the betweenness values of the publications' authors such that publications of central authors are ranked on top.

However, two particular problems emerge from that model. One is the conceptual problem of author name ambiguity (homonymy, synonymy) in bibliographic databases. In particular the potential homonymy of names may misrepresent the true social structure of a scientific community. The other problem is the computation effort needed for calculating betweenness in large networks that may bother, in case of long computation times, the retrieval process and finally user acceptance.

## IV. IMPLEMENTATION

All proposed models were implemented in an online information system sowiport to demonstrate the general feasibility of the three approaches. The prototype uses those models as particular search stratagems (Bates 1990) to enhance retrieval quality. The open source search server Solr is used as the basic retrieval engine which provides a standard term frequency based ranking mechanism (TF-IDF). All three models work as retrieval add-ons on-the-fly during the retrieval process.

The Bradfordizing re-ranking model is implemented as a Solr plugin which orders all search results with an ISSN number such that the journal with the highest ISSN count gets the top position in the result set, the second journal the next position, and so forth. The numerical TF-IDF ranking value of each journal paper in the result set is then multiplied with the frequency count of the respective journal. The result of this computation is taken for re-ranking such that core journal publications are ranked on the top.

The author centrality based re-ranking model computes a co-authorship network on the basis of the result set retrieved for a query, according to the co-authorships appearing in the result set documents. For each node in the graph betweenness is measured, and each document in the result set is assigned a relevance value given by the maximum betweenness value of its authors. Single authored publications are captured by this method if their authors appear in the graph due to other publications they have published in co-authorship. Thus, just publications from pure single fighters are ignored by this procedure. The result set is then re-ranked by the centrality value of the documents' authors such that publications of central authors appear on the top of the list.

## V. SUMMARY

In our longer paper (Mutschke et al. 2011) we could show empirically that the structural models of science can be used to improve retrieval quality. The other way around, the IR experiment turned out that to the same extent to which science models contribute to IR (in a positive as well as negative sense), science-model driven IR might contribute to a better understanding of different conceptualizations of science (role of journals, authors and language in scientific discourses). Recall and precision values of retrieval results obtained by science model oriented search and ranking techniques seem to provide important indicators for the adequacy of science models in representing and predicting structural phenomena in science.

A further point that might be interesting from the perspective of science modeling is the degree of acceptance of science models as retrieval methods by the users of a scholarly IR system. The degree to which scientists are willing to use those models for finding what they are looking for (as particular search stratagems, as proposed by Bates 2002) are further relevant indicators for the degree to which the models intuitively meet the real research process. Thus, the major contributions of IR to science modeling might be to measure the expressiveness of existing science models and to generate novel models from the perspective of IR. In addition, the application and utilization of science model enhanced public retrieval systems can probably be a vehicle to better explain and communicate science and science models to a broader audience in the sense of public understanding of science.

However, a lot of research effort needs to be done to make more progress in coupling science modeling with IR. We see this paper as a first step in this area. The major challenge that we see here is to consider also the dynamic mechanisms which form the structures and activities in question and their relationships to dynamic features in scholarly information retrieval

ACKNOWLEDGMENT

We would like to express our grateful thanks to Andrea Scharnhorst for her valuable comments and collaboration. She was also one of the co-organizers of the recent workshop at ISSI "Combining Bibliometrics and Information Retrieval". We thank Philipp Schaer and Thomas Lüke for the main implementation work in the project IRM I and IRM II. IRM I was funded by DFG, grant no. INST 658/6-1. IRM II was funded by DFG, under grant no. SU 647/5-2.